\documentclass[twocolumn,showpacs,preprintnumbers,amsmath,amssymb,aps,prc]{revtex4-2}
\usepackage[colorlinks=true,linkcolor=blue,citecolor=blue,urlcolor=blue]{hyperref}
\usepackage{mathtools}
\usepackage{amsmath}
\usepackage{subfigure}
\usepackage{bm}
\usepackage[dvipsnames]{xcolor}

\begin{document}

\renewcommand{\thetable}{\arabic{table}}

\hypersetup{
    colorlinks=true,
    linkcolor=blue,
    citecolor=blue,
    urlcolor=blue
}

\title{Physics-informed interpretable forms for dipole resonances and quantities relevant for astrophysics}
\author{Gourab Banerjee$^{1,2}$, Jhilam Sadhukhan$^{2,3}$, Debasish Mondal$^{2,3}$, B.K.Agrawal$^{1}$ ,Chandrani Sen$^{2,3}$, S.Mukhopadhyay$^{2,3}$\\
$^1$Saha Institute of Nuclear Physics, 1/AF Bidhannagar, Kolkata 700064, India, \\
$^2$Homi Bhabha National Institute, Mumbai 400094, India, \\
$^3$Variable Energy Cyclotron Centre, 1/AF Bidhannagar, Kolkata 700064, India}

\begin{abstract}
We have employed Bayesian inference to extract the centroid energies and widths from the isovector giant dipole resonance (IVGDR) data for the IVGDR built on the ground states of nuclei. These IVGDR properties were modeled by explicitly incorporating the structural effects like isospin asymmetry, nuclear deformation and shell correction. The resulting parametrizations reproduce the IVGDR systematics with remarkable precision for both spherical and axially deformed nuclei. Subsequently, the extracted IVGDR parameters were used to constrain the coefficients of the nuclear symmetry energy, which are found to be in good agreement with recent measurements of nuclear dipole polarizability. Furthermore, these symmetry-energy coefficients were employed to predict the dipole polarizability within different self-consistent mean-field models and to investigate its correlation with the neutron-skin thickness. The present findings provide valuable input for large-scale calculations of radiative capture processes relevant to nuclear astrophysics and for improving our understanding of astrophysical scenarios under extreme conditions.
\end{abstract}

\maketitle

%{\it Introduction:} 
\section{Introduction}
The isovector giant dipole resonance (IVGDR) is a prominent collective excitation mode that provides crucial insight into nuclear structure and dynamics. In particular, it serves as a sensitive probe of the nuclear symmetry energy and shear viscosity—quantities that play key roles in understanding a wide range of astrophysical phenomena such as the structure of neutron stars and dynamics of neutron star merger~\cite{Lim:2023dbk}. In the macroscopic picture, the IVGDR is interpreted as an out-of-phase oscillation of proton and neutron fluids within the nucleus~\cite{harakeh2001giant,Bracco:2007zz,Chakrabarty:2016kna,Gaardhoje:1992tz,Santonocito:2020clw,Snover:1986dg,Santonocito:2018pom,Bracco:2019gza,Alhassid:1999rdi}. Microscopically, it can be described as a coherent superposition of one-particle–one-hole (1p–1h) excitations~\cite{JSpeth_1981,Speth:1991myq}. The properties of the IVGDR are studied experimentally using the photoabsorption reaction, which shows an enhanced Lorentzian strength in the 10–25 MeV region characterized by the peak energy ($E_G$), the width ($\Gamma_G$), and the strength ($S_G$). Accurate determination of these quantities is crucial for understanding the IVGDR decay mechanism and for determining the photon strength function, a key quantity in modelling the nucleosynthesis process. 

Further, the photoabsorption cross section is directly related to the nuclear dipole polarizability, which provides strong constraints on the magnitude, density dependence, and slope of the nuclear symmetry energy~\cite{Roca-Maza:2015eza}. Another important isovector property of a nucleus is the neutron-skin thickness. This observable has attracted considerable attention, both experimentally and theoretically, because it is closely connected to the density dependence of the symmetry energy around nuclear saturation density~\cite{Centelles:2008vu,Roca-Maza:2011qcr,Agrawal:2012pq}. Notably, the neutron-skin thickness can also be inferred from IVGDR properties, owing to the established correlation between dipole responses and the neutron–proton density asymmetry.

Both microscopic and phenomenological approaches have been extensively employed to describe the properties of IVGDR, particularly $E_G$ and $\Gamma_G$. Early interpretations of IVGDR were based on two idealized models of nuclear collective motion. In the Goldhaber–Teller (GT) model~\cite{Goldhaber:1948zza}, protons and neutrons are treated as incompressible fluids oscillating against each other through mutual displacement~\cite{Nuclearmodel}. This picture leads to a resonance energy that scales as $E_G\propto A^{-1/6}$, which happens to be more suitable for light nuclei~\cite{ring1980nuclear}. By contrast, the Steinwedel–Jensen (SJ) model~\cite{SJ} assumes a nucleus of fixed spherical geometry in which the resonance arises from density waves of protons and neutrons oscillating out of phase within the nuclear volume. In this case, the centroid energy follows the scaling $E_G\propto A^{-1/3}$. Both models conserve global isospin symmetry and are therefore most suitable for nuclei with $N = Z$. Experimental systematics reveal that real nuclei display a mixed behavior, lying between these two limiting descriptions. This indicates that the IVGDR dynamics reflect a more intricate interplay of surface and volume contributions than either model alone can fully capture.

A deeper understanding of IVGDR dynamics requires a fully microscopic quantum many-body treatment. Time-dependent Hartree-Fock (TDHF) theory~\cite{Simenel:2018euu} and its pairing-inclusive extension, time-dependent Hartree-Fock-Bogoliubov (TDHFB) theory~\cite{Simenel:2012xq}, are well suited for this purpose. The random phase approximation (RPA)~\cite{Goeke:1982qm,Bertsch} and, for open-shell nuclei, its quasiparticle extension (QRPA), are also frequently used to compute IVGDR properties. While microscopic models generally predict the centroid energy $E_G$ with good accuracy, they often struggle to reproduce the width $\Gamma_G$~\cite{Bai:2021bqw}. Moreover, classical hydrodynamic models—with their simple analytical expressions—remain more practical for large-scale applications, where full microscopic calculations are computationally demanding. 

Experimentally, photoabsorption cross sections for nuclei in their ground states are typically analyzed by fitting Lorentzian functions to the measured spectra using the least-square method, thereby extracting the centroid energy $E_G$, width $\Gamma_G$, and the IVGDR strength $S_G$~\cite{Plujko:2018uum}. More recently, machine-learning approaches—such as artificial neural networks and multitask neural networks—have been applied to predict IVGDR parameters directly from experimental data. While these data-driven models can achieve high predictive accuracy within the domain of their training sets, they often lack transparency regarding the explicit role of underlying nuclear properties. These considerations motivate the development of improved analytical formulations for IVGDR observables that retain clear physical interpretability while ensuring broad applicability across the nuclear chart. In the present work, we adopt a Bayesian inference framework to generalize and optimize the functional forms of $E_G$, $\Gamma_G$, and related quantities. In particular, we extend the analysis to extract constraints on the nuclear symmetry energy, dipole polarizability, and neutron-skin thickness, with special emphasis on their utility in large-scale nuclear structure calculations and astrophysical applications.

The paper is organized as follows. The details of the theoretical model are given in Sec.~\ref{sec:model}. Subsequently, the results are elaborated in Sec.~\ref{sec:result}, followed by the conclusions in Sec.~\ref{sec:conclusion}.  

\section{Theoretical Model}\label{sec:model}
\subsection{IVGDR parameters}
In the energy range of 10–25~MeV, the total photoabsorption cross-section of a nucleus is typically dominated by IVGDR, and the corresponding spectra are usually fitted with a simplified Lorentzian (SLO) distribution~\cite{brink1955ph,Axel:1962zz} to extract the experimental IVGDR parameters. According to this model, the $\gamma$ resonance cross section for a nucleus of triaxial ground state, in general, is given by 
\begin{equation}
   \sigma_{abs}(E_{\gamma})=\frac{2\sigma_{TRK}}{\pi}\sum_{i=1}^{3}\frac{S_{G_i}\Gamma_{G_i} E^2_{\gamma}}{(E^2_{\gamma}-E^2_{G_i})^2+E^2_{\gamma} \Gamma^2_{G_i}},  
\label{eq-cross}
\end{equation}
where $E_\gamma$ denotes the $\gamma$-ray energy, and $\sigma_{TRK}=\frac{60 N Z}{A}$ mb MeV, as follows from the Thomas–Reiche–Kuhn (TRK) dipole sum rule. Here, $E_{G_i}$, $\Gamma_{G_i}$, and $S_{G_i}$ are the IVGDR parameters along the 
$i$-th principal axis of the triaxial nucleus. Let us collectively denote them by $\bm{\chi}_G\equiv\{\chi_{G_i},i=1,2,3\}$ with $\chi_{G_i}\equiv\{E_{G_i},\Gamma_{G_i}, S_{G_i}\}$. For spherical nuclei $\chi_{G_1}=\chi_{G_2}=\chi_{G_3}\equiv \chi_G^o$. For axially deformed nuclei, $\chi_{G_1}\ne \chi_{G_2}=\chi_{G_3}$, where, for convenience, we denote the parameters along the symmetry axis by $\chi_{G}^{\parallel}$ and those perpendicular to it by $\chi_G^\perp$. The parameters $\bm\chi_G$ are determined by fitting the measured IVGDR spectra, and the resulting values are taken as the experimental ones.

Bayesian inference provides an elegant and statistically rigorous framework for simultaneously optimizing the parameters $\bm\chi_G$ in Eq.~(\ref{eq-cross}) from the measured photoabsorption spectra~\cite{Sen:2025mti}. Bayes’ theorem reads~\cite{10.1111/j.1467-985X.2014.12096_1.x}
\begin{equation}
P(X,\bm\chi_G|D)=\frac{\mathcal{L}(D|X,\bm\chi_G) P(\bm\chi_G)}{\mathcal{Z}},
\label{eq-bayes}
\end{equation}
where $P(\bm\chi_G)$ is the prior probability distribution of the parameters, and $P(X,\bm\chi_G|D)$ is the posterior distribution. Here, $D=\{d_1,d_2,..,d_N\}$
denotes the experimental data set with $N$ measured points at locations 
$X=\{x_1,x_2...,x_N\}$. The variable $X$ may represent a single nuclear property (e.g., $A$, $Z$, or $E_\gamma$) or a combination thereof. The likelihood function is expressed as $\mathcal{L}(D|X,\bm\chi_G)=\prod_{i=1}^{N} \mathcal{L}(d_i|x_i,\bm\chi_G)$ with individual terms
\begin{equation}
\mathcal{L}(d_i|x_i,\bm\chi_G)=\frac{1}{\sqrt{2\pi\sigma_i^2}}\exp{\left\{\frac{-\left(d_i-M(x_i;\bm\chi_G)\right)^2}{2\sigma_i^2}\right\}},
\label{eq-like}
\end{equation}
where $M(x_i;\bm\chi_G)$ is the model prediction for the quantity measured at $x_i$ and $\sigma_i$ is the experimental uncertainty for the corresponding experimental data $d_i$. The normalization factor $\mathcal{Z}= \sum_{\bm\chi_G}\mathcal{L}(D|X,\bm\chi_G) P(\bm\chi_G)$ is known as the Bayesian evidence. The Bayesian inference naturally quantifies correlations among model parameters. Let $\theta$ and $\phi$ be two parameters with posterior distributions $P(\theta)$ and $P(\phi)$, and means $\bar{\theta}$ and $\bar{\phi}$. The correlation coefficient between them is~\cite{casella2002statistical}
\begin{equation}
    r[\theta,\phi]=\frac{\sum_{i=1}^{N}\{P(\theta_i) - \bar{\theta}\}\{P(\phi_i) - \bar{\phi}\}}{\sqrt{\sum_{i=1}^{N}\{P(\theta_i) - \bar{\theta}\}^2} \sqrt{\sum_{i=1}^{N}\{P(\phi_i) - \bar{\phi}\}^2}}.
\label{eq-corr}
\end{equation}
The bounds of this coefficient are: $ -1\le r[\theta,\phi]\le 1$. 

For spherical nuclei, the phenomenological expression traditionally used for the IVGDR centroid energy is~\cite{Myers:1977zz}
\begin{equation}
    E_G^o(A)=a_0A^{-\frac{1}{3}}+a_1A^{-\frac{1}{6}},
\label{eq-eg1}
\end{equation}
a combination of the GT and SJ model scalings. The standard values $a_0=31.2$~MeV and $a_1=20.6$~MeV were determined in Ref. ~\cite{Gaardhoje:1992tz} from early experimental data. However, Eq.~(~\ref{eq-eg1}) is ideally valid only for spherical, isospin-symmetric ($N=Z$) nuclei. In general, the centroid energy should depend on the isospin asymmetry $I=|N-Z|/A$. Hence, we propose
\begin{equation}
 \begin{aligned}
     E_{G}^o(A,I)=(a_0 A^{-\frac{1}{3}} + a_1 A^{-\frac{1}{6}})(1- a_I I^2),
 \end{aligned}
 \label{eq-eg2}
\end{equation}
where the isospin dependence is adopted following a similar dependence in the liquid–drop model of binding energy~\cite{Myers:1966zz, Myers:1982hs}. The above expression introduces three parameters, $\{a_0, a_1, a_I\}$, all of which need to be inferred from the measured data using Bayesian analysis.

For deformed nuclei, the centroid energy along the axis of symmetry is expressed as~\cite{Capote:2009qqf},
\begin{equation} \label{eq-eg-parallel}
E_{G}^\parallel(A,I,\beta_2)=E_{G}^o(A,I)\left(1+\sqrt{\frac{5}{4\pi}}\beta_2\right)^{-1},
\end{equation}
where $\beta_2$ represents the Bohr's deformation parameter that quantifies the quadrupole deformation of axially symmetric nuclei. Although $\beta_2$ can be taken from existing literature, for consistency, we obtained it from the measured splitting in the IVGDR strength distribution, i.e., from the ratio $E_G^\perp/E_G^\parallel$. Precisely, $\beta_2$ is related to the ratio ($\eta$) of the lengths of the major and minor axes of an axial nucleus as~\cite{Okamoto1956,Okamoto:1958zz,DANOS195823,Berman:1975tt}
\begin{equation}
    \beta_2=\frac{8\pi}{15}\sqrt{\frac{5}{16\pi}} \left(\eta^2 -1\right) \eta^{-\frac{2}{3}}.
\label{eq-beta}
\end{equation}
Further, according to the hydrodynamic model, $\eta$ can be obtained from the measured centroid energies as~\cite{Okamoto1956,Okamoto:1958zz,DANOS195823,Berman:1975tt}
\begin{equation}
    \eta=\frac{1}{0.911}\left(\frac{E_G^\perp}{E_G^\parallel}-0.089\right),
\label{eq-eta}
\end{equation}
yielding $\eta=1$, {\it viz-a-viz} $\beta_2=0$, for spherical nuclei.

For a spherical nucleus of mass $A$, charge $Z$, and radius $R_0$, the surface-dissipation model~\cite{Bush:1991jtp} gives the IVGDR width as
\begin{equation}
    \Gamma_G^o=k_s\frac{ZN}{A} \frac{Z^{-1} \bar{v}_p + N^{-1} \bar{v}_n}{R_0},
\label{eq-gamma-spherical}
\end{equation}
where $\bar{v}_n$ and $\bar{v}_p$ are the average neutron and proton velocities, respectively, and these are determined from their Fermi energies. In general, the average velocity is given by $\bar{v}=\frac{3}{4}v_F$, where $v_F=(2\epsilon_F/m)^{1/2}$ is the Fermi velocity and $\epsilon_F$ denotes the Fermi energy. The parameter $k_s$ $(<1)$ accounts for nonclassical corrections that reduce the dissipation. For axially deformed nuclei, the widths along and perpendicular to the symmetry axis follow the power law~\cite{DANOS1964113}:
\begin{equation}
\Gamma_{G}^{\parallel(\perp)}=\Gamma_G^o\left(\frac{E_{G}^{\parallel(\perp)}}{E_G^o}\right)^{\delta_{\parallel(\perp)}},
\label{eq-gamma-deformed}
\end{equation}
where $\delta_\parallel$ and $\delta_\perp$ are determined by fitting the experimental data, and $E_G^o$ and $\Gamma_G^o$ are the centroid energy and width corresponding to a hypothetical spherical shape of the same nucleus, respectively. Employing the relation between $E_{G}^\parallel$ and $E_G^o$ given in Eq.~(\ref{eq-eg-parallel}), we get 
\begin{equation}\label{eq-gammag-parallel}
    \Gamma_G^\parallel=\Gamma_G^o\left(1+\sqrt{\frac{5}{4\pi}}\beta_2\right)^{-\delta_\parallel}.
\end{equation}
Further, application of Eq.~(\ref{eq-eta}) leads to
\begin{equation}\label{eq-gammag-perp}
    \Gamma_G^\perp=\Gamma_G^o\left\{\left(1+\sqrt{\frac{5}{4\pi}}\beta_2\right)(0.911\eta+0.089)\right\}^{\delta_\perp}.
\end{equation}
Often, a simplified parametrized form for the IVGDR width in terms of the IVGDR centroid energy is adopted in the literature~\cite{Capote:2009qqf}. It is expressed as
\begin{equation} \label{eq-gamma-all} \Gamma_G^{o,\parallel,\perp}=c\left(E_G^{o,\parallel,\perp}\right)^\delta,
\end{equation}
where the parameters take the values $c = 0.026$ and $\delta = 1.9$ for spherical nuclei, while $\delta = 1.91$ is used for deformed nuclei.

\subsection{Astrophysical relevant quantities}
In IVGDR, the neutron and proton fluids oscillate out of phase, establishing a direct connection with the nuclear symmetry energy. Within the hydrodynamic model of Danos~\cite{DANOS195823}, the symmetry energy coefficient $K$ of the Weizsäcker semi-empirical mass formula can be extracted from the IVGDR centroid energy and width. Using the measured $E_G^\parallel$ and $\Gamma_G^\parallel$, one finds the experimental $K$ as~\cite{Berman:1975tt}
\begin{equation}
\begin{aligned}
    K=9.935\times10^{-4}\frac{A^{8/3}}{NZ}\left\{\frac{\left(E_{G}^\parallel\right)^2}{1-\left(\frac{\Gamma_{G}^\parallel}{2E_{G}^\parallel}\right)^2}\right\}\\
    \times \Biggr[\frac{\eta^{\frac{4}{3}}}{(1+0.01860\epsilon-0.03314\epsilon^2)^2}\Biggr],
\end{aligned}
\label{eq-symm}
\end{equation}
where $\epsilon$ is the eccentricity of the axially deformed nucleus, and is related to $\eta$. For spherical nuclei, $E_G^\parallel\rightarrow E_G^o$, $\Gamma_G^\parallel\rightarrow\Gamma_G^o$, and the term in square brackets reduces to unity. Alternatively, $K$ may be parametrized as a function of mass number as~\cite{Berman:1975tt}
\begin{equation}
K=K_v-K_s A^{-\frac{1}{3}},
\label{eq-kfact}
\end{equation}
where $K_v$ and $K_s$ are known as the volume and surface symmetry energy coefficients, respectively, and these can be optimized with the help of {\it measured} $K$ in Eq.~(\ref{eq-symm}).

Another IVGDR-related observable of considerable astrophysical relevance is the nuclear dipole polarizability $\alpha_D$. It can be obtained by integrating the measured photoabsorption cross-section as 
\begin{equation}
 \alpha_D=\frac{\hbar c}{2 \pi^2 }\int_{0}^{\infty} \frac{\sigma_{abs}(E_{\gamma})}{E_{\gamma}^2} \,dE_{\gamma}.
\label{eq-polar-abs}
\end{equation}
Alternatively, within semi-classical approaches—particularly the leptodermous expansion developed by Lipparini and Stringari—the dipole polarizability can be expressed as~\cite{LIPPARINI1989103}
\begin{equation}
    \alpha_D=\frac{e^2}{12}\frac{A\langle r^2\rangle}{K_v}\left(1+\frac{5}{3}\frac{K_s}{K_v}A^{-\frac{1}{3}}\right),
\label{eq-polar2}
\end{equation}
where $\langle r^2\rangle$ denotes the mean-square nuclear radius averaged over the nuclear volume. For a spherical nucleus with a uniform mass distribution, the mean-square radius reduces to $\langle r^2\rangle=(3/5)R_0^2$. A more realistic estimate of $\langle r^2\rangle$, however, can be obtained within self-consistent mean-field models employing effective nucleon–nucleon interactions. The use of a mean-field framework is particularly advantageous because it provides simultaneous access to several bulk isovector properties of nuclei. In particular, mean-field models can yield~\cite{Dutra:2012mb} the nuclear symmetry energy $J$ obtained at the saturation density $\rho_0$, as well as the neutron skin thickness, defined as
\begin{equation}
    r_\text{skin}=\langle r_n^2\rangle^{1/2}-\langle r_p^2\rangle^{1/2}, 
\end{equation}
the difference in the root-mean-square radii of the neutron and proton fluids. It is now well established~\cite{Roca-Maza:2013mla} that the product $\alpha_DJ$ exhibits a strong correlation with $r_\text{skin}$. Consequently, this correlation offers a powerful means to estimate $r_\text{skin}$ by combining the mean-field prediction of $\alpha_D$ in Eq.~(\ref{eq-polar2}) with the optimum $K_v$ and $K_s$ values from  Eq.~(\ref{eq-kfact}). Such an approach provides a consistent bridge between IVGDR observables and fundamental isovector properties of nuclear matter.

Alternatively, $\alpha_D$ can also be evaluated within the framework of the droplet model, which incorporates essential information from the nuclear equation of state. The corresponding expression for $\alpha_D$ reads~\cite{MYERS1974186,Roca-Maza:2013mla}
\begin{equation}
    \alpha_D=\frac{\pi e^2}{54}\frac{A\langle r^2\rangle}{J}\left(1+\frac{5}{3}\frac{L}{J}\epsilon_A\right),
\label{eq-polar1}
\end{equation}
where
\begin{equation}
  \epsilon_A=\frac{\rho_0 - \rho_A}{3 \rho_0}\quad\text{with}\quad \rho_A=\frac{\rho_0}{1+dA^{-1/3}}.
\label{eq-rhoA}
\end{equation}
Here, $\rho_0$ denotes the nuclear saturation density, and the parameter $d$ is determined by imposing the condition $\rho_A(A=208)$=0.1 fm$^{-3}$. In Eq.~(\ref{eq-polar1}), $L$ represents the slope of the symmetry energy at saturation density $\rho_0$. It would be interesting to examine how the results are modified when $\alpha_D$ from Eq.~(\ref{eq-polar1}) is employed instead of that from Eq.~(\ref{eq-polar2}). 
%%%%%%%%%%%%%%%%%%%%%%%%%%%%%  FIG 1  %%%%%%%%%%%%%%%%%%%%%%%%%%%%%%%%%%%%%%
\begin{figure}
    \centering
    \includegraphics[width=\linewidth]{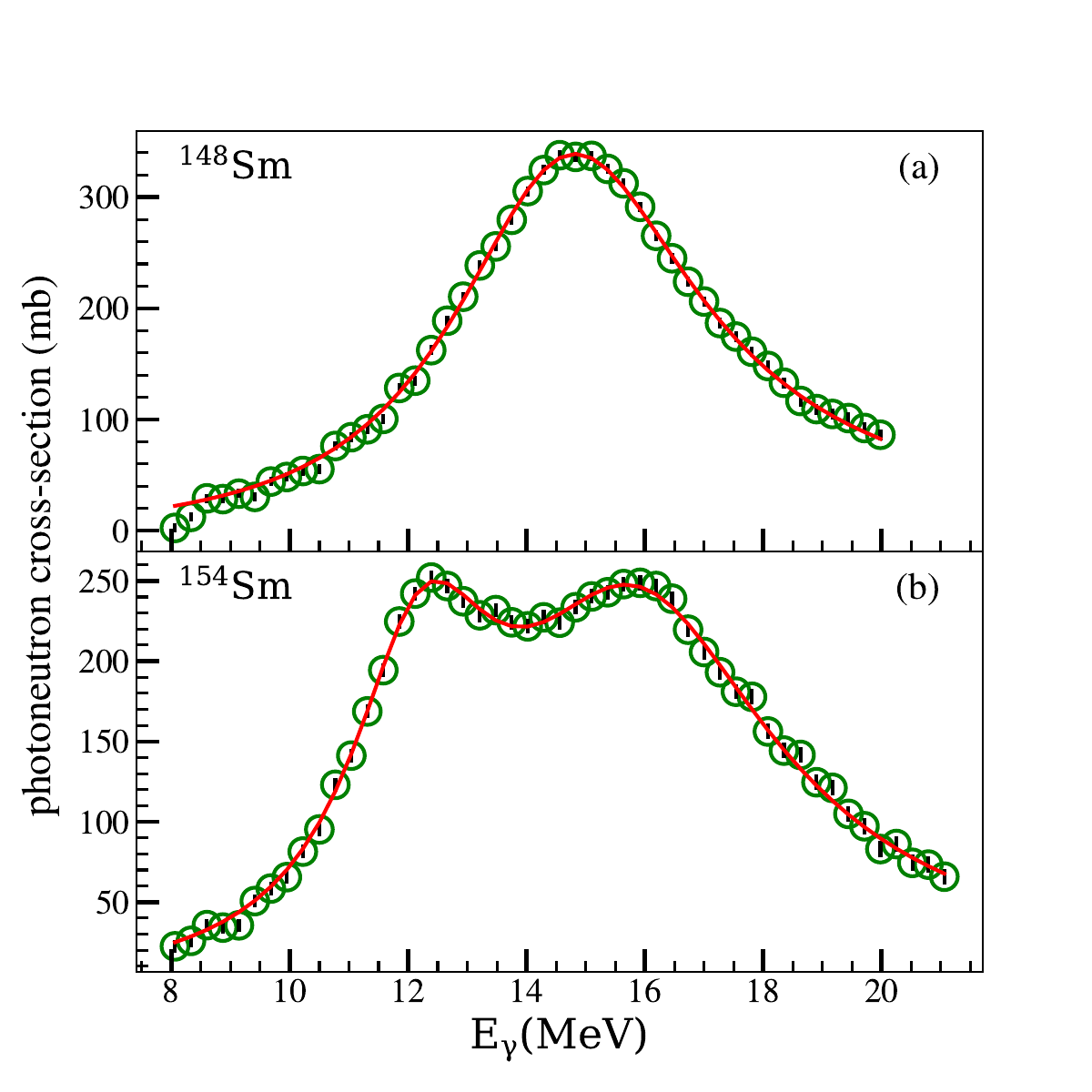}
    \caption{(Color online) Measured photoneutron cross-sections (symbols) of spherical $^{148}$Sm (top panel) and deformed $^{154}$Sm (bottom panel) are compared with the calculated spectra (solid lines) from the SLO model in Eq.~(\ref{eq-cross}), where the IVGDR parameters are estimated using the Bayesian analysis.}
    \label{plot-cross}
\end{figure}
%%%%%%%%%%%%%%%%%%%%%%%%%%%%%%%%%%%%%%%%%%%%%%%%%%%%%%%%%%%%%%%%%%%%%%%%%
\section{Results}\label{sec:result}
The Bayesian inference framework is employed to extract the experimental IVGDR parameters, $\bm{\chi}_G$, as defined within the standard Lorentzian (SLO) model of Eq.~(\ref{eq-cross}). The analysis uses the measured $\gamma$-ray spectra for 81 spherical and 60 axially deformed nuclei, compiled in the EXFOR database~\cite{EXFOR} or in the references therein. These spectra constitute the data set $D$ for the Bayesian evaluation described in Eq.~(\ref{eq-bayes}). A uniform prior distribution for $P(\bm{\chi}_G)$ is adopted throughout this optimization, as well as in all subsequent Bayesian analyses presented in this work. Fig.~\ref{plot-cross} presents a typical comparison between the experimental spectra and the fitted spectra obtained using the SLO model, where the IVGDR parameters are estimated from the Bayesian approach.

\subsection{IVGDR centroid energy}
\begin{figure}
    \centering
    \includegraphics[width=0.9\linewidth]{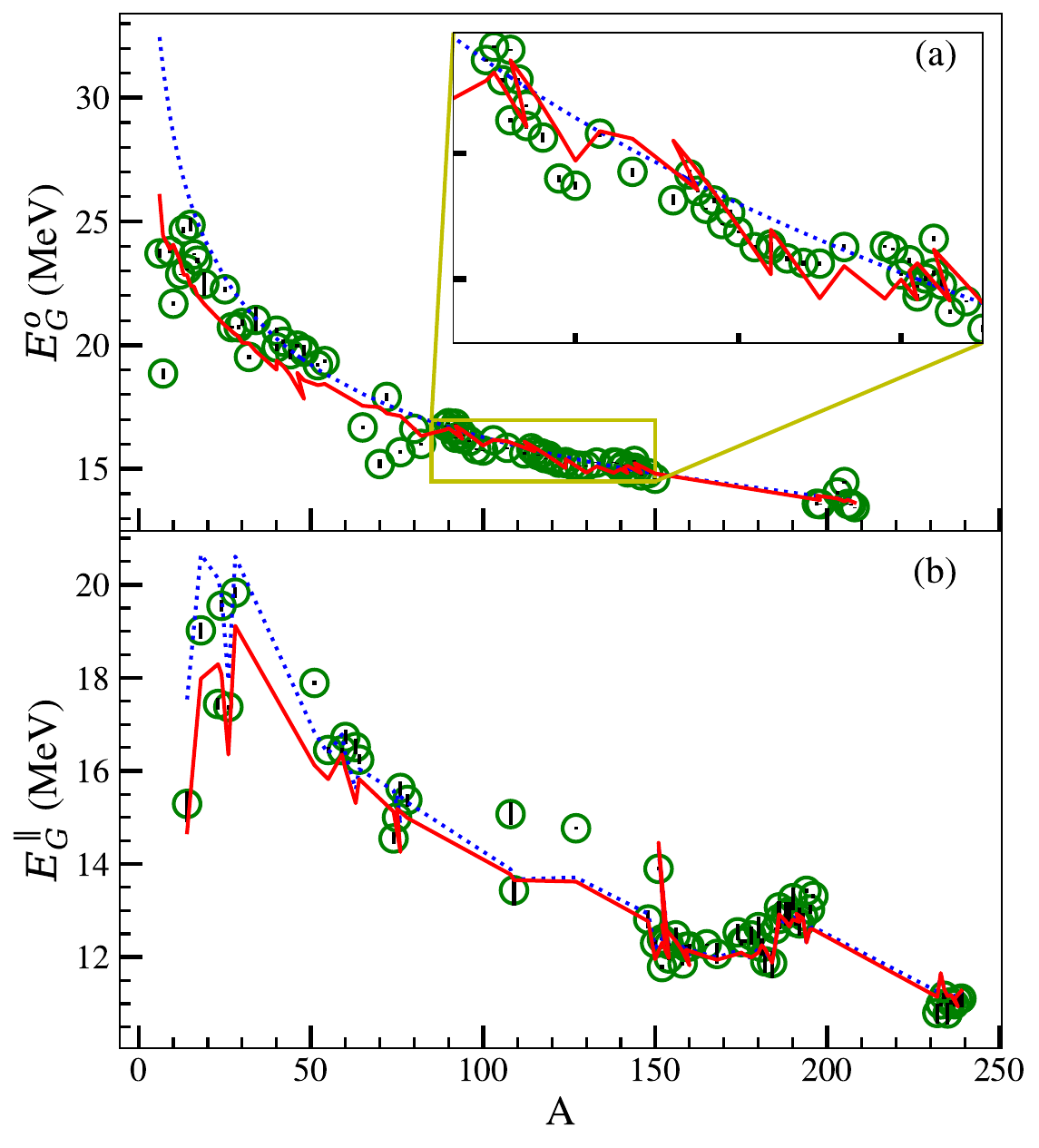}
    \caption{(Color online) The experimental (symbols) IVGDR centroid energies obtained from the Bayesian analysis for spherical (top panel) and axially deformed (bottom panel) nuclei are compared with the values calculated (solid lines) from Eq.~(\ref{eq-eg2}) and Eq.~(\ref{eq-eg-parallel}), respectively. The $E_G^{o,\parallel}$ without the isospin factor (see text) are shown by dotted lines.}
    \label{plot-eg}
\end{figure}

The extracted $E_G^{o,\parallel}$ values, identified as the experimental IVGDR centroid energies, are displayed in Fig.~\ref{plot-eg} to enable a more detailed analysis. The ratio $E_G^\perp/E_G^\parallel$ is substituted into Eq.~(\ref{eq-eta}) to evaluate the asymmetry parameter $\eta$, which, in turn, is used in Eq.~(\ref{eq-beta}) to determine the quadrupole deformation parameter $\beta_2$. The resulting $\beta_2$ values exhibit excellent agreement with those inferred from traditional $\gamma$-ray spectroscopy of low-lying excited states~\cite{Raman:2001nnq}, affirming the reliability of the present extraction technique.
%%%%%%%%%%%%%%%%%%%%%%%%%%%%%  FIG 2   %%%%%%%%%%%%%%%%%%%%%%%%%%%%%%%%%%
%%%%%%%%%%%%%%%%%%%%%%%%%%%%%%%%%%%%%%%%%%%%%%%%%%%%%%%%%%%%%%%%%%%%%%%%%%

The parameters entering Eq.~(\ref{eq-eg2}) are further constrained through an additional Bayesian analysis by utilizing the experimental values of $E_G^{o}$ together with their associated $1\sigma$ uncertainties. The resulting predictions of $E_G^{o}$ from Eq.~(\ref{eq-eg2}) and of $E_G^{\parallel}$ from Eq.~(\ref{eq-eg-parallel}) (using the extracted $\beta_2$ values) are compared with the corresponding experimental data in Fig.~\ref{plot-eg}. For completeness, results obtained from the existing phenomenological expression without the isospin-dependent term (Eq.~(\ref{eq-eg1})) are also displayed in the same figure. As expected, both formulations reproduce the overall mass dependence of the centroid energy, with only a minor discrepancy between them in the low-mass region below $A = 50$. Notably, for spherical nuclei, the proposed expression successfully captures the subtle isospin dependence of the centroid energies. This demonstrates that the newly introduced analytical form in Eq.~(\ref{eq-eg2}) offers a more reliable and precise framework for estimating IVGDR centroid energies over a wide range of nuclear masses.

\begin{figure}
    \centering
    \includegraphics[width=0.9\linewidth]{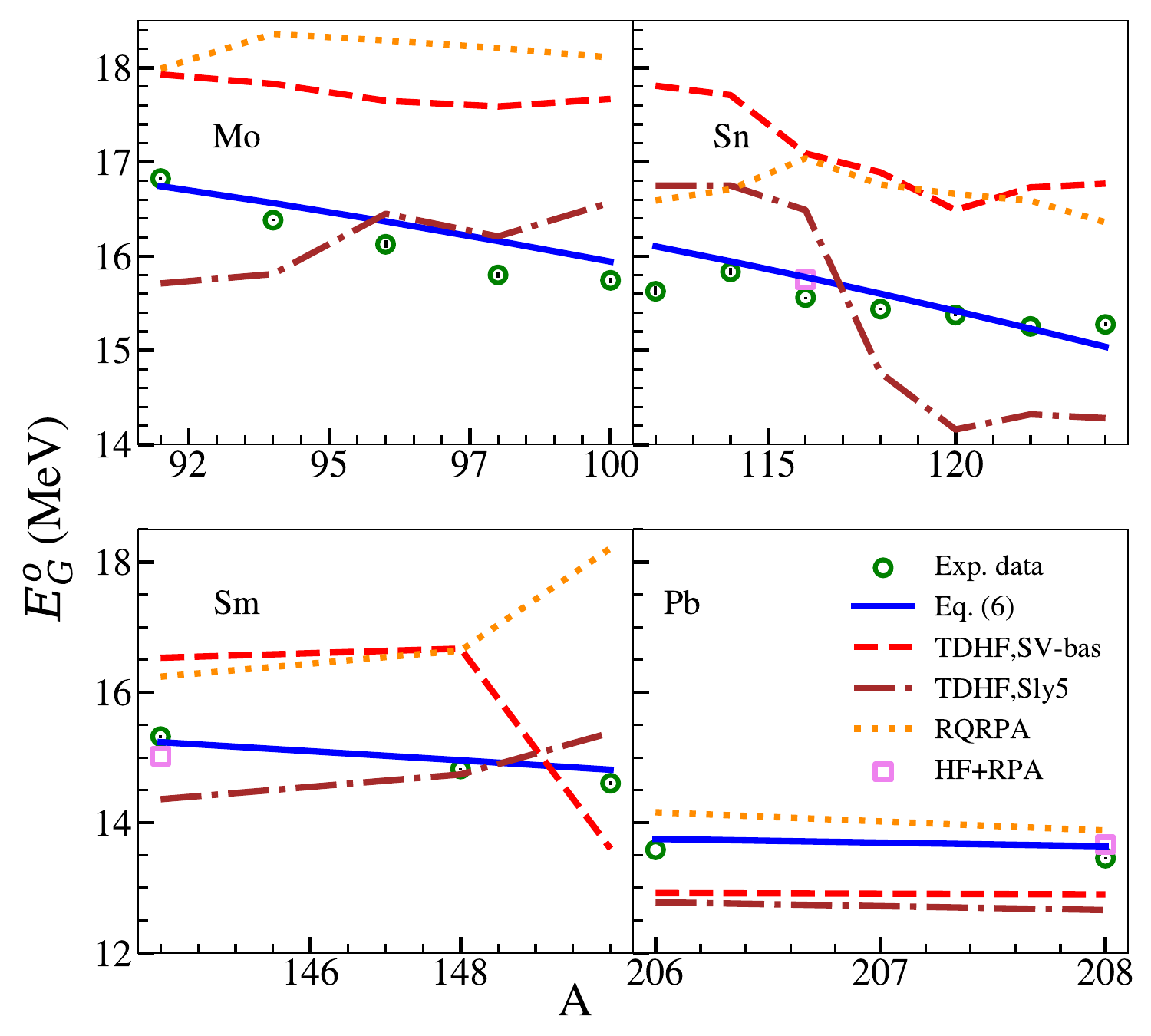}
    \caption{(Color online) IVGDR centroid energies calculated from Eq.~(\ref{eq-eg2}) are compared with theoretical results from TDHF, RQRPA and the experimental data. A few data points from HF+RPA calculation in~\cite{Sil:2006sh} are included.}
    \label{plot-microEg}
\end{figure}

We have further examined the above proposition by comparing the IVGDR centroid energies predicted by Eq.~(\ref{eq-eg2}) with results from microscopic calculations. Four isotopic chains of spherical nuclei spanning different mass regions (Mo, Sn, Sm, and Pb) are considered. The microscopic calculations are based on two theoretical approaches: the time-dependent Hartree--Fock (TDHF) theory~\cite{Abhishek:2024inq} and the relativistic quasiparticle random phase approximation (RQRPA)~\cite{Yuksel:2019dnp} employing the relativistic density-dependent point-coupling (DDPCX) interaction. The TDHF calculations are performed using two Skyrme energy density functionals, SV-bas~\cite{Klupfel:2008af} and SLy5~\cite{Chabanat:1997un}. These results, together with the predictions of Eq.~(\ref{eq-eg2}), are compared with the available experimental data in Fig.~\ref{plot-microEg}. For selected nuclei, centroid energies obtained from Hartree--Fock plus random phase approximation (HF+RPA) calculations~\cite{Sil:2006sh} are also included. The microscopic calculations reproduce the overall experimental trend reasonably well. However, the TDHF results obtained with the SV-bas and SLy5 parametrizations exhibit noticeable differences for most isotopic chains, except Pb, indicating the sensitivity of $E_G^o$ to the underlying energy density functional. The TDHF predictions based on the SV-bas interaction are generally consistent with the RQRPA results. In addition, the HF+RPA centroid energies for $^{116}$Sn, $^{144}$Sm, and $^{208}$Pb closely agree with both the predictions of Eq.~(\ref{eq-eg2}) and the corresponding experimental data. Overall, Eq.~(\ref{eq-eg2}) reproduces the measured centroid energies remarkably well across the considered nuclei and correctly captures the isospin dependence, demonstrating its robustness and suitability for regions where experimental data are unavailable.

\begin{figure}
    \centering
    \includegraphics[width=0.9\linewidth]{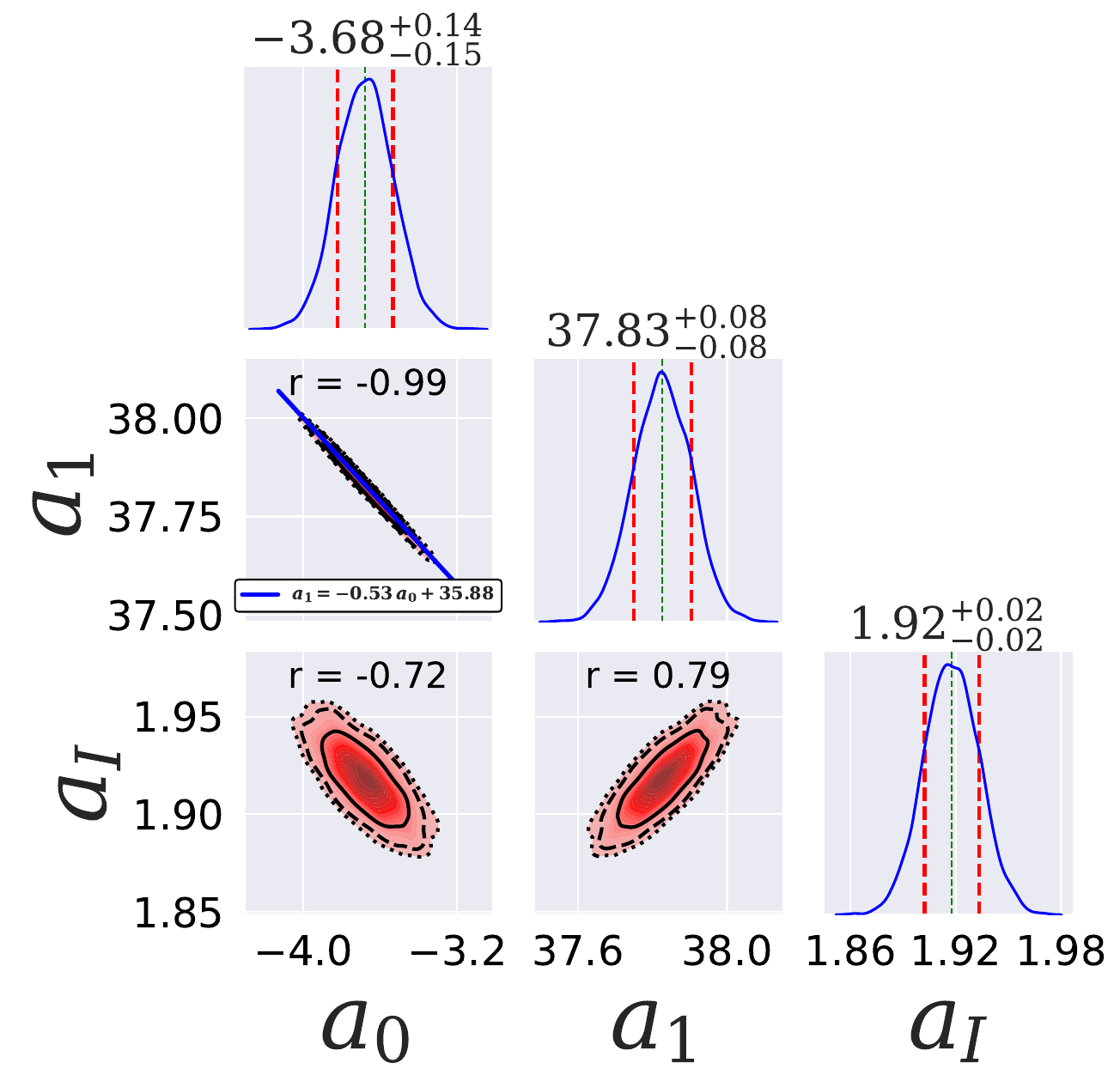}
    \caption{(Color online) The correlations (contours) among various parameters of Eq.~(\ref{eq-eg2}) as indicated. For each parameter, the probability distribution (solid blue line) is presented with indications of most-probable value (green vertical line) and 1$\sigma$ error (red dotted lines).}
    \label{plot-cor}
\end{figure}

The correlations among the parameters $a_0$, $a_1$, and $a_I$ in Eq.~(\ref{eq-eg2}), along with their respective posterior probability distributions, are shown in Fig.~\ref{plot-cor}. The most probable values of $a_0$ and $a_1$ are found to deviate from those implied by Eq.~(\ref{eq-eg1}). These shifts arise from two primary factors: the inclusion of additional data points with $A < 40$, which were not considered in the earlier analysis, and the incorporation of the isospin-dependent term in Eq.~(\ref{eq-eg2}). Interestingly, Fig.~\ref{plot-cor} exhibits a pronounced anti-correlation between $a_0$ and $a_1$ (correlation coefficient $r \approx -0.99$), indicating a strong mutual dependence between the contributions proportional to $A^{-1/3}$ and $A^{-1/6}$. Although the GT and SJ models were originally formulated based on different fluid-dynamical assumptions, this result suggests that the $A^{-1/3}$ and $A^{-1/6}$ dependencies may, in fact, originate from a common underlying physical mechanism.
%%%%%%%%%%%%%%%%%%%%%%%%  FIG 3    %%%%%%%%%%%%%%%%%%%%%%%%%%%%%%%%%%%%%%%
%%%%%%%%%%%%%%%%%%%%%%%%%%%%%%%%%%%%%%%%%%%%%%%%%%%%%%%%%%%%%%%%%%%%%%%%
%%%%%%%%%%%%%%%%%%%%%%%%%%  FIG 4   %%%%%%%%%%%%%%%%%%%%%%%%%%%%%%%%%%%%
%%%%%%%%%%%%%%%%%%%%%%%%%%%%%%%%%%%%%%%%%%%%%%%%%%%%%%%%%%%%%%%%%%%%%%%%%
\subsection{IVGDR width}
\begin{figure}
    \centering
    \includegraphics[width=0.9\linewidth]{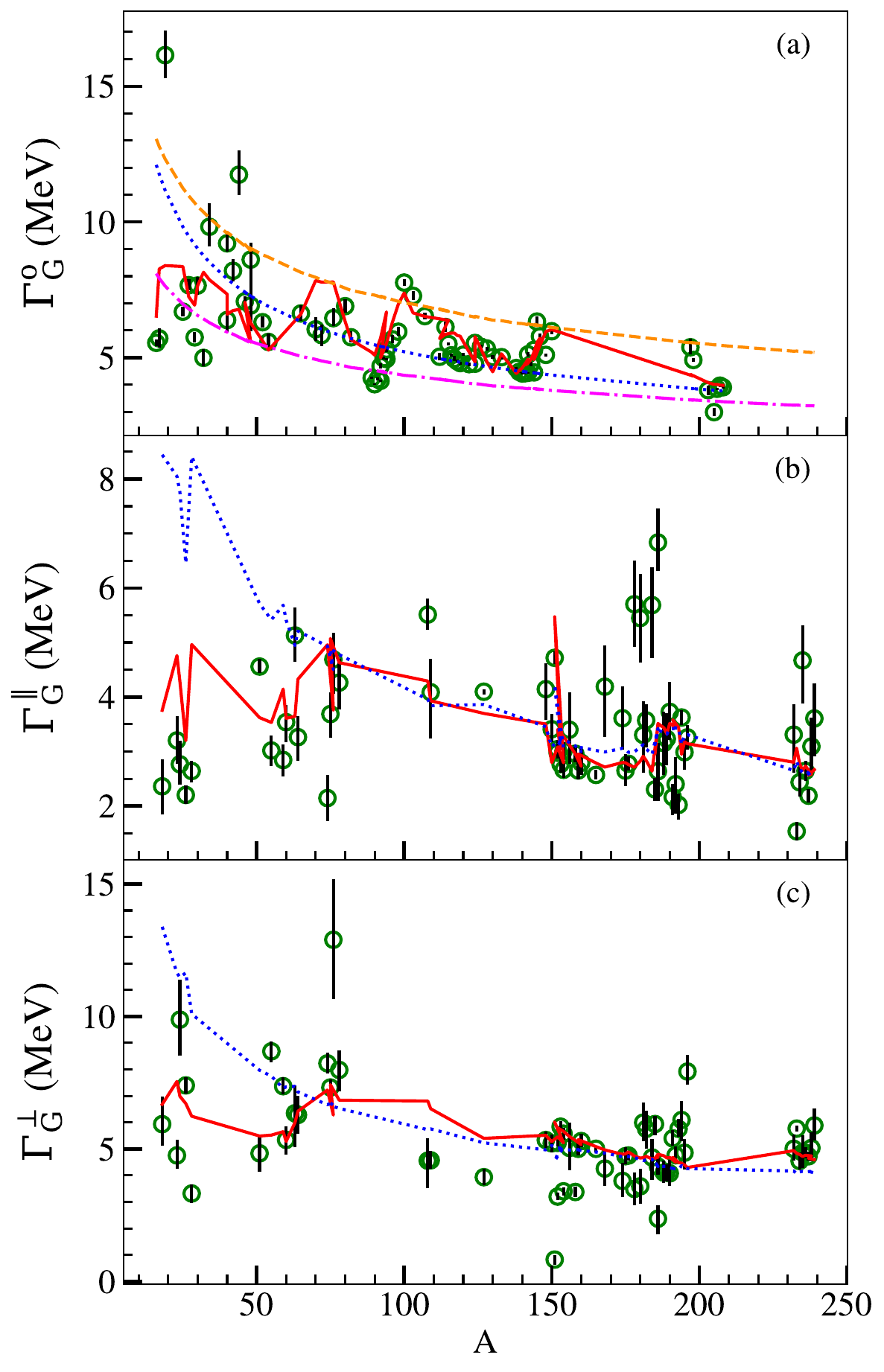}
    \caption{(Color online) Experimental (symbols) IVGDR widths for spherical (top panel) and axially deformed (middle and bottom panels) nuclei are compared with the predictions (solid lines): (1) $\Gamma_G^o$ from Eq.~(\ref{eq-gamma-spherical}) with $k_s$ in Eq.~(\ref{eq-ks}), (2) $\Gamma_G^\parallel$ ($\Gamma_G^\perp$) from Eq.~(\ref{eq-gammag-parallel}) (Eq.~(\ref{eq-gammag-perp})) with $\Gamma_G^o$ from Eq.~(\ref{eq-gamma-spherical}), and $\delta_\parallel=3.66$ ($\delta_\perp=-1.06$). $\Gamma_G^{o}$ (top panel) for $k_s=1$ (dashed line) and $k_s=0.62$ (dash-dotted line), and  $\Gamma_G^{o,\parallel,\perp}$ (all panels) from Eq.~(\ref{eq-gamma-all}) (dotted line)  are shown for reference.
    }
    \label{plot-gammag}
\end{figure}

Fig.~\ref{plot-gammag} presents the systematic behavior of the experimental IVGDR widths, $\Gamma_G^{o,\parallel,\perp}$, which, on average, show a gradual decrease with increasing $A$. This average trend characterizes nuclear bulk dynamics and, for spherical nuclei, can be reproduced using Eq. (\ref{eq-gamma-spherical}) with a constant value of $k_s$ and Eq.~(\ref{eq-gamma-all}) as well. The top panel of Fig.~\ref{plot-gammag} compares $\Gamma_G^o$ with predictions for two extreme values of $k_s$ (1.0 and 0.62), as quoted in the literature~\cite{Bush:1991jtp}; the experimental widths lie within these bounds. Nonetheless, significant local fluctuations are observed, which influence the smooth global $A$ dependence.

\begin{figure}
    \centering
    \includegraphics[width=\linewidth]{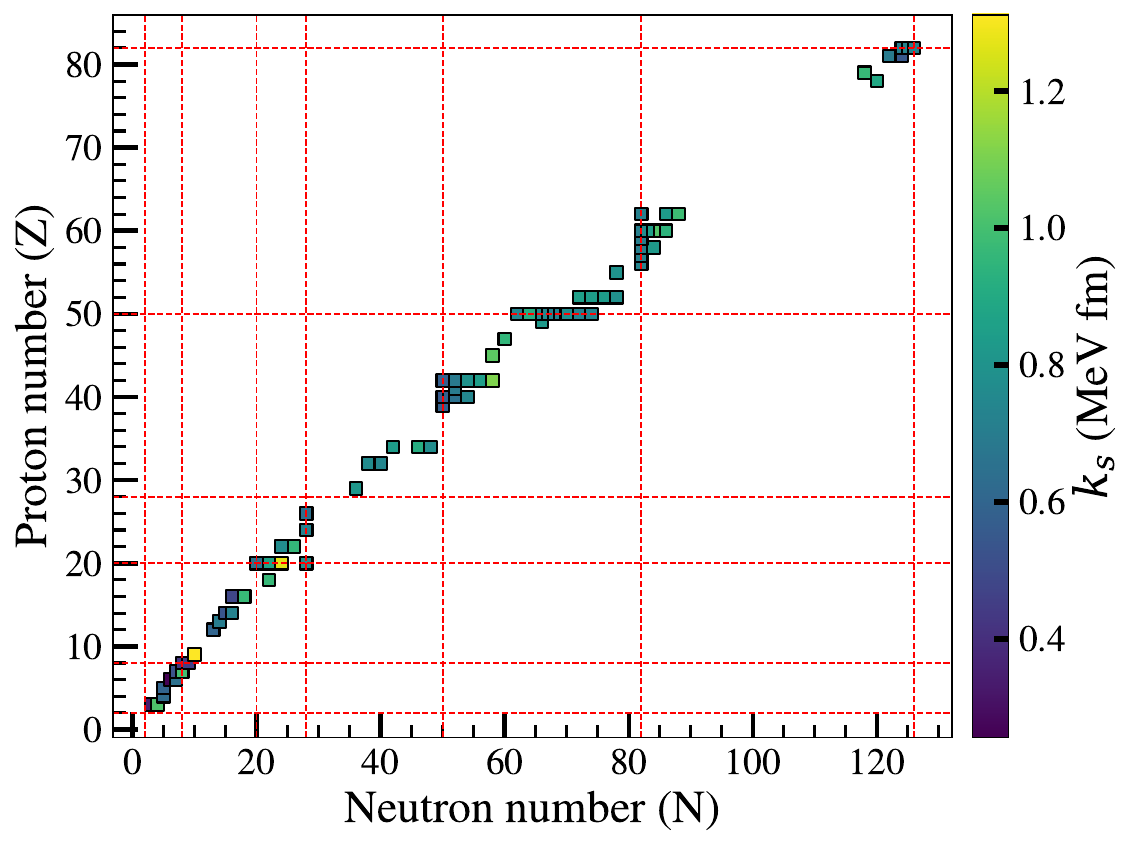}
    \caption{(Color online) Optimum values of $k_s$ for spherical nuclei considered in this work. Neutron and proton magic numbers are marked with dotted lines.}
    \label{plot-ks}
\end{figure}

To incorporate these structural effects, we introduce a refined description of the damping coefficient $k_s$. We first determine $k_s$ by fitting the experimental $\Gamma_G^o$ for each spherical nucleus. The extracted $k_s$ values are shown in Fig.~\ref{plot-ks},
%%%%%%%%%%%%%%%%%%%%%%%%%%  FIG 5   %%%%%%%%%%%%%%%%%%%%%%%%%%%%%%%%%%%%
%%%%%%%%%%%%%%%%%%%%%%%%%%%%%%%%%%%%%%%%%%%%%%%%%%%%%%%%%%%%%%%%%%%%%%%%%
revealing that $k_s$ is systematically suppressed near neutron or proton magic numbers and increases toward unity in mid-shell regions, except for some nuclei below $A=30$. This behavior reflects the competition between dissipative dynamics and microscopic shell effects, the latter being dominant at closed shells. To capture this dependence quantitatively, we examined $k_s/A^{1/3}$ that leads to the following functional form:
\begin{equation}
   k_s(\delta W,A)=\{0.05\exp{(0.21 \delta W)} + 0.12\}A^{1/3},
\label{eq-ks}
\end{equation}
where $\delta W$ denotes the ground-state shell-correction energy as given in~\cite{Moller:1993ed}. The coefficients in Eq. (\ref{eq-ks}) were 
%%%%%%%%%%%%%%%%%%%%%%%%  FIG 6   %%%%%%%%%%%%%%%%%%%%%%%%%%%%%%%%%%
%%%%%%%%%%%%%%%%%%%%%%%%%%%%%%%%%%%%%%%%%%%%%%%%%%%%%%%%%%%%%%%%%%%
extracted through a fit to the $k_s$ values in Fig.~\ref{plot-ks}. As illustrated in Fig.~\ref{plot-gammag}, substituting this expression into Eq.~(\ref{eq-gamma-spherical}) leads to a marked improvement in the reproduction of the experimental $\Gamma_G^{o}$ values, clearly outperforming predictions obtained with a constant $k_s$. In earlier works~\cite{10.1007/978-94-009-7099-1_167,Goriely:1998utv}, ground-state shell corrections were incorporated into the IVGDR width by explicitly including inputs from independent nuclear structure studies. In contrast, the present approach provides a more compact and straightforward formulation, making it particularly convenient for large-scale applications. 

For axially deformed nuclei, the IVGDR widths are evaluated using Eq.~(\ref{eq-gammag-parallel}) and Eq.~(\ref{eq-gammag-perp}). In an earlier investigation~\cite{Bush:1991jtp}, the exponents were taken as $\delta^{\parallel}=\delta^{\perp}=1.6$, assuming a constant $k_s=0.62$ in $\Gamma_G^{o}$. In the present work, we have reexamined the values of $\delta^{\parallel}$ and $\delta^{\perp}$ by employing the mass-dependent form of $k_s$ given in Eq.~(\ref{eq-ks}). This analysis yields $\delta^{\parallel}=3.66$ and $\delta^{\perp}=-1.06$. Adopting a single exponent $\delta=\delta^{\parallel}=\delta^{\perp}$ proves inadequate to reproduce the complete experimental systematics. As demonstrated in Fig.~\ref{plot-gammag}, the proposed formulation closely tracks the experimental data across most of the nuclear chart. In particular, a significantly improved agreement is achieved in the low-mass region ($A \lesssim 50$) compared with the predictions of Eq.~(\ref{eq-gamma-all}). Interestingly, the improved isospin-dependent expressions for the IVGDR centroid energy and width could play an important role in determining the photon strength function for unstable nuclei that are not experimentally accessible. It will have crucial implications for nuclear reaction modelling relevant to astrophysical nucleosynthesis processes. 

\subsection{Astrophysical relevant properties}
\begin{figure}
    \centering
\includegraphics[width=0.9\linewidth]{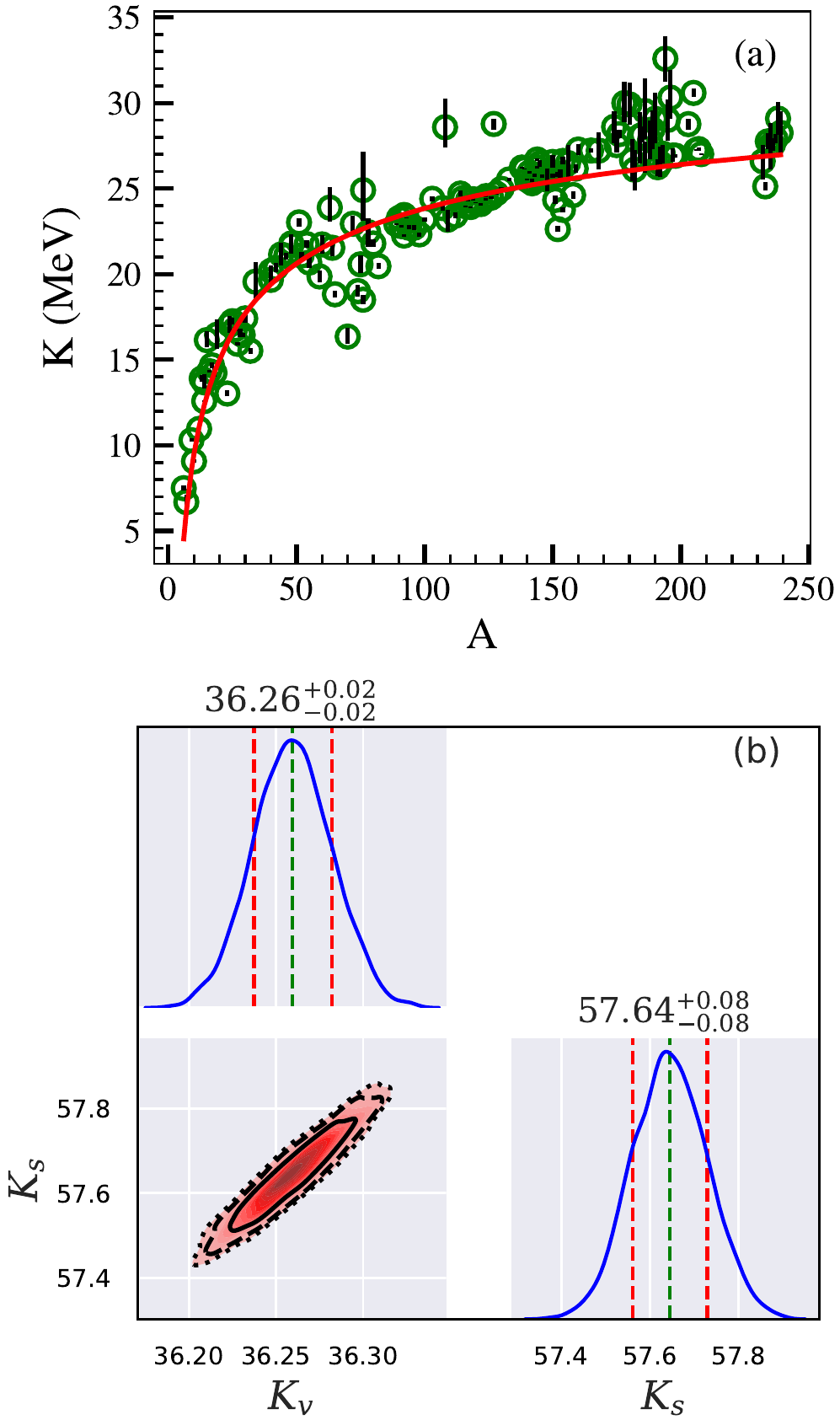}
\caption{(Color online) (a) Experimental data (symbols) of nuclear symmetry energy coefficient ($K$) for spherical and deformed nuclei are compared with the predictions (solid line) from Eq.~(\ref{eq-kfact}). (b) Corresponding distributions and correlation of symmetry energy coefficients $K_v$ and $K_s$ (Eq.~(\ref{eq-kfact})) obtained from Bayesian inference using the experimental data in (a).}
    \label{plot-symm}
\end{figure}

The experimental values of the symmetry energy coefficient $K$ for axially deformed (spherical) nuclei are extracted by substituting the measured $E_G^{\parallel}$ ($E_G^{o}$) and $\Gamma_G^{\parallel}$ ($\Gamma_G^{o}$) into Eq.~(\ref{eq-symm}). The resulting values are displayed in Fig.~\ref{plot-symm} as a function of $A$, and are subsequently employed to optimize the parameters $K_v$ and $K_s$ in Eq.~(\ref{eq-kfact}). As evident from Fig.~\ref{plot-symm}, the Bayesian analysis produces smooth, unimodal posterior probability distributions for both $K_v$ and $K_s$. The most probable values are found to be $K_v = 36.26^{+0.02}_{-0.02}$ and $K_s = 57.64^{+0.08}_{-0.08}$, in good agreement with those reported earlier~\cite{Jiang:2012zzb}. In addition, a moderate correlation between $K_v$ and $K_s$ is observed, highlighting their interdependence within the optimization procedure.
%%%%%%%%%%%%%%%%%%%%%%%%%%%  FIG 7   %%%%%%%%%%%%%%%%%%%%%%%%%%%%%%%%%%%%
%%%%%%%%%%%%%%%%%%%%%%%%%%%%%%%%%%%%%%%%%%%%%%%%%%%%%%%%%%%%%%%%%%%%%%%%%%
\begin{table}[]
    \centering
    \begin{tabular}{|c|c|c|c|}
    \hline
        EoS & L & J & $\rho_0$  \\
    \hline
        SKM$^*$ &  45.78  &  30.03 &  0.160    \\
        \hline
        SLY4 & 45.94  &  32.00 &  0.160    \\
        \hline
        SLY5 & 48.15  &  32.01 &  0.161    \\
        \hline
        SIII & 9.91  &  28.16 &  0.145    \\
        \hline
        SKP & 19.68  &  30.00  & 0.163   \\
        \hline
        SKO & 79.14  &  31.97 &  0.160     \\
        \hline
        SKI3 & 100.53  &  34.83 &  0.158      \\
        \hline
        SKX & 33.18  &  31.10 &  0.155       \\
        \hline
    \end{tabular}
    \caption{The saturation properties for different parametrizations of the Skyrme EDFs.}
    \label{table_edf}
\end{table}
%%%%%%%%%%%%%%%%%%%%%%%%%%  FIG 8   %%%%%%%%%%%%%%%%%%%%%%%%%%%%%%%%%%%%%%
\begin{figure}
    \centering
    \includegraphics[width=\linewidth]{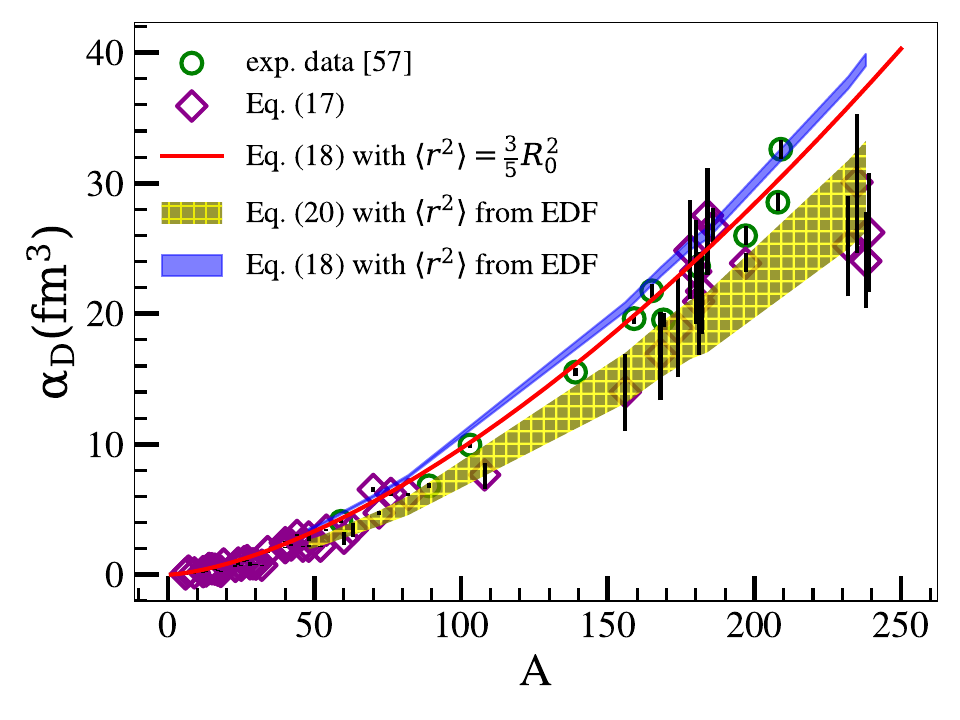}
    \caption{(Color online) Experimental dipole polarizability ($\alpha_D$) from the literature~\cite{Goriely:2020hnk} (open circles) and that extracted from photoabsorption data using Eq.~(\ref{eq-polar-abs}) (open diamonds) are compared with theoretical predictions: Eq.~(\ref{eq-polar2}) with $\langle r^2 \rangle = \frac{3}{5}R_0^2$ (solid red line); Eq.~(\ref{eq-polar2}) with $\langle r^2 \rangle$ from different EDFs (blue band); and Eq.~(\ref{eq-polar1}) with $\langle r^2 \rangle$ from different EDFs (yellow band).}
    \label{plot-polar}
\end{figure} 
%%%%%%%%%%%%%%%%%%%%%%%%%%%%%%%%%%%%%%%%%%%%%%%%%%%%%%%%%%%%%%%%%%%%%%%%%%

Subsequently, we compare the measured $\alpha_D$ values with the predictions obtained from Eq.~(\ref{eq-polar2}) and Eq.~(\ref{eq-polar1}). The experimental $\alpha_D$ values are taken from two different sources: (1) the recent photoneutron cross-section data reported in Ref.~\cite{Goriely:2020hnk}, and (2) those extracted from the photoabsorption spectra used in the present analysis, following Eq.~(\ref{eq-polar-abs}). For dataset (2), a total of 52 photoabsorption spectra are considered, including both spherical and deformed nuclei. In this case, the Bayesian-optimized $\sigma_{abs}$ [Eq.~(\ref{eq-cross})] is incorporated into Eq.~(\ref{eq-polar-abs}) over a broad range of $E_\gamma$, ensuring that the calculated $\alpha_D$ becomes insensitive to the choice of the integration limit. The extracted $\alpha_D$ values are shown in Fig.~\ref{plot-polar}. Overall, the two datasets exhibit good agreement, except for some deviations around $A\approx160$ and $A\approx240$. This discrepancy appears to be associated with the intrinsic trend of the IVGDR data, as a similar feature is also observed for $K$ in Fig.~\ref{plot-symm}. Additional data in the mass region $200\lesssim A\lesssim 240$ would be valuable for clarifying this behavior.

%%%%%%%%%%%%%%%%%%%%%%%%%%%%%  FIG 9  %%%%%%%%%%%%%%%%%%%%%%%%%%%%%%%%%%%%%
\begin{figure}
    \centering
    \includegraphics[width=\linewidth]{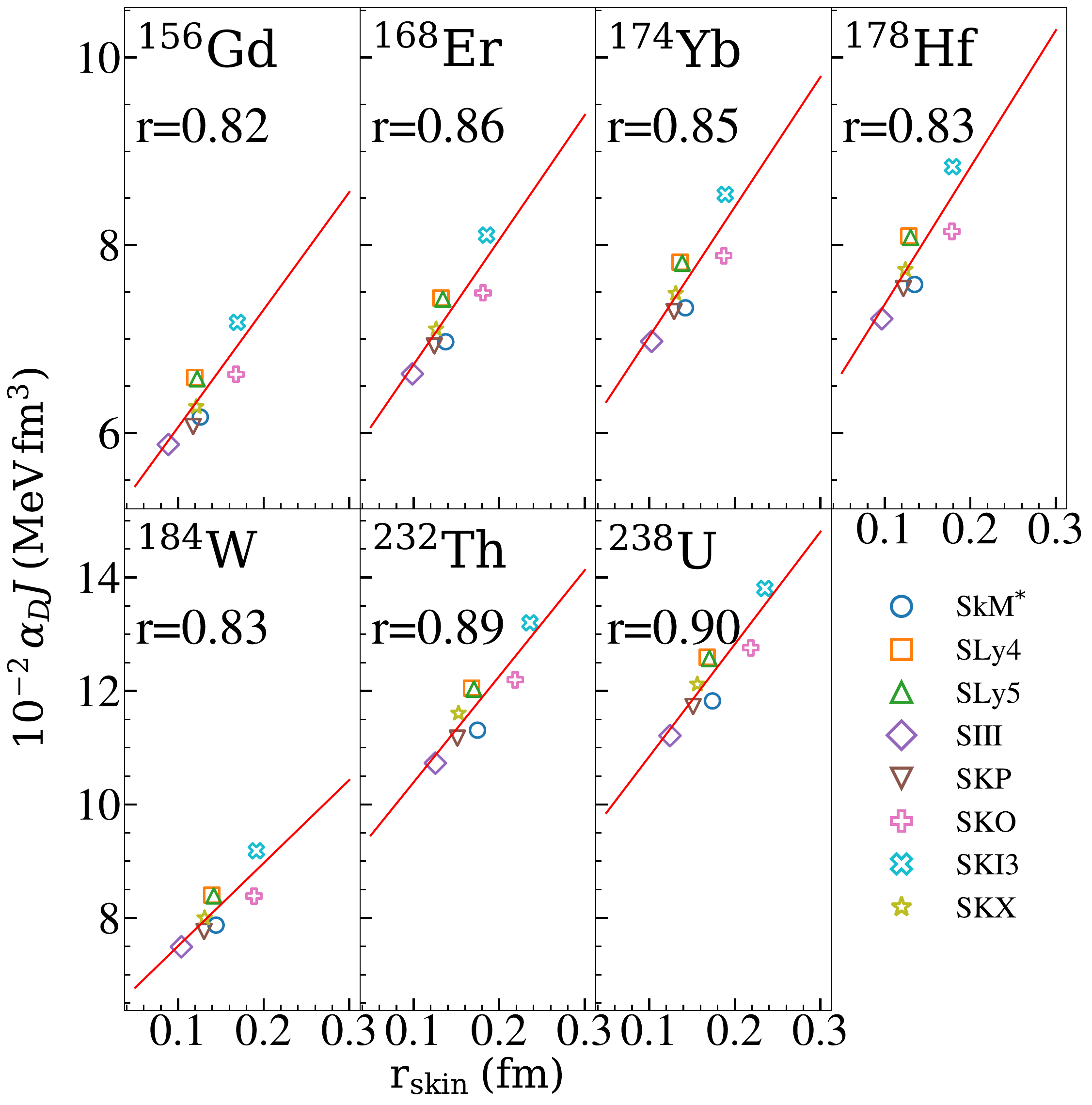}
    \caption{(Color online) Correlation of the product $\alpha_D J$ with neutron skin thickness ($r_\text{skin}$) depending on the choice of the Skyrme parametrization (symbols) employed to obtain $\alpha_D$ from Eq.~(\ref{eq-polar2}). The selected nucleus and the correlation coefficient $r$ (Eq.~(\ref{eq-corr})) are mentioned in each panel. The best-fit straight lines are shown as solid lines.}
    \label{plot-polar2}
\end{figure} 
%%%%%%%%%%%%%%%%%%%%%%%%%%%%%%%%%%%%%%%%%%%%%%%%%%%%%%%%%%%%%%%%%%%%%%%%%%
%%%%%%%%%%%%%%%%%%%%%%%%%%%  FIG 10   %%%%%%%%%%%%%%%%%%%%%%%%%%%%%%%%%%%%%%
\begin{figure}
    \centering
    \includegraphics[width=\linewidth]{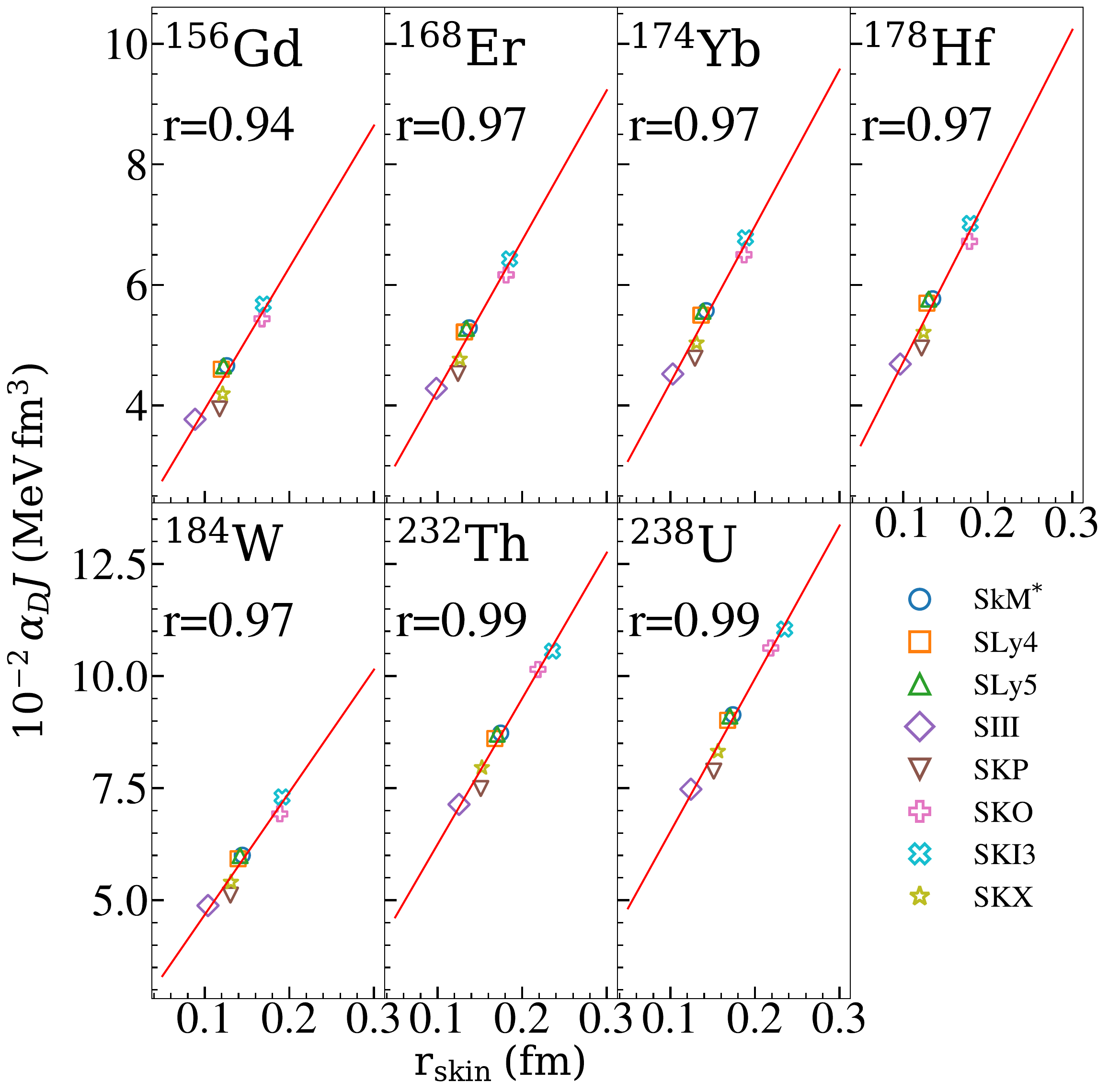}
    \caption{Same as Fig.~\ref{plot-polar2}, but now $\alpha_D$ is calculated from Eq.~(\ref{eq-polar1}).}
    \label{plot-polar2-new}
\end{figure} 
%%%%%%%%%%%%%%%%%%%%%%%%%%%%%%%%%%%%%%%%%%%%%%%%%%%%%%%%%%%%%%%%%%%%%%%%%%%

For the theoretical estimation of $\alpha_D$ from Eq.~(\ref{eq-polar2}), the symmetry energy coefficients $K_v$ and $K_s$ extracted from the IVGDR analysis are substituted into Eq.~(\ref{eq-polar2}). Two different prescriptions are adopted for the mean-square radius $\langle r^2 \rangle$ in this equation. In the first approach, $\langle r^2 \rangle$ is calculated within the nuclear energy density functional (EDF) framework using the HFODD code (version v2.49t)~\cite{Schunck:2011qp}. In this calculation, the Hartree–Fock–Bogoliubov equation is solved using eight different Skyrme EDF parametrizations—$\mathrm{SkM}^*$, SLy4, SLy5, SIII, SKP, SKO, SKI3, and SKX in the particle–hole channel—together with a density-dependent mixed pairing interaction~\cite{Dobaczewski:2002mt} in the particle–particle channel. In the second approach, a simplified estimate $\langle r^2 \rangle = \frac{3}{5}R_0^2$ is employed, with $R_0 = 1.2A^{1/3}$. The calculated $\alpha_D$ values are compared with the experimental data in Fig.~\ref{plot-polar}. All the EDFs considered in the present study yield very similar $\alpha_D$ values, which are also close to those obtained using the simplified radius prescription. Moreover, our predictions—derived directly from the analysis of the IVGDR spectra—show good agreement with the measured data, particularly the photoneutron data reported in Ref.~\cite{Goriely:2020hnk}. In addition, Eq.~(\ref{eq-polar2}) appears to provide an upper bound to the available experimental data. 

Further, $\alpha_D$ is independently evaluated from Eq.~(\ref{eq-polar1}) using the nuclear equation-of-state (EOS) parameters $J$, $L$, $\rho_0$, and $\langle r^2 \rangle$ corresponding to the EDFs mentioned above. The adopted EOS parameters are listed in Table~\ref{table_edf}. As illustrated in Fig.~\ref{plot-polar}, the resulting $\alpha_D$ values exhibit a broader uncertainty band arising from the choice of EDF, and this spread increases with the mass number. Furthermore, Eq.~(\ref{eq-polar1}) appears to reproduce the lower bound of the experimental data, thereby complementing the $\alpha_D$ values predicted by Eq.~(\ref{eq-polar2}).
%%%%%%%%%%%%%%%%%%%%%%  FIG 11   %%%%%%%%%%%%%%%%%%%%%%%%%%%%%%%%%%%%%%%%%%%%
\begin{figure}
    \centering
    \includegraphics[width=0.85\linewidth]{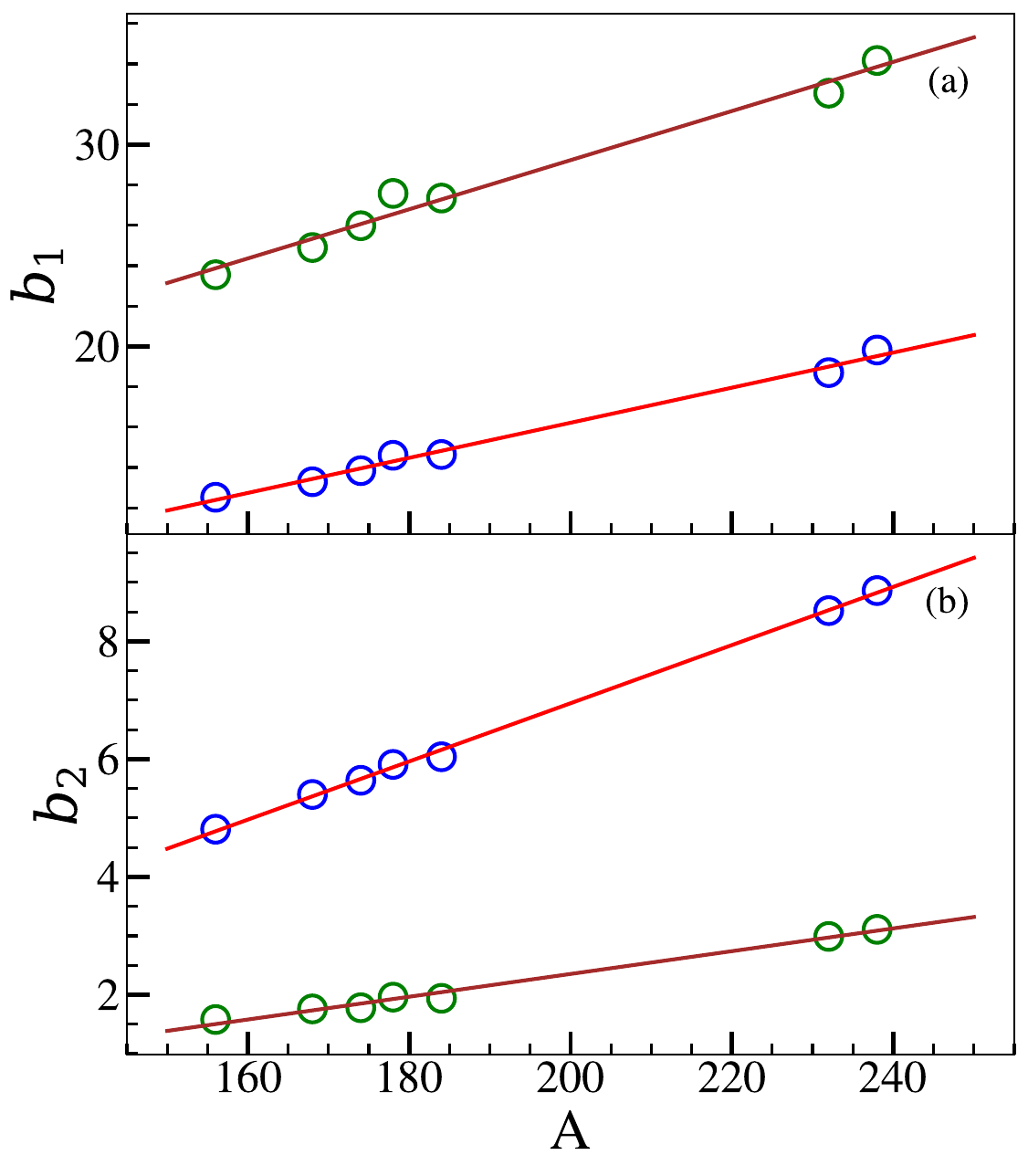}
    \caption{The plot for the fitting coefficients $b_1$ $\&$ $b_2$ for $10^{-2}\alpha_D J$ vs r$_{skin}$ for the diiferent massive system.}
    \label{plot-polar3}
\end{figure} 
%%%%%%%%%%%%%%%%%%%%%%%%%%%%%%%%%%%%%%%%%%%%%%%%%%%%%%%%%%%%%%%%%%%%%%%%%%

Next, we investigate how the product $\alpha_DJ$ correlates with the neutron-skin thickness $r_\text{skin}$ for different choices of EDF. For consistency, $r_\text{skin}$ is calculated using the same EDF employed to determine $J$ and $\langle r^2\rangle$ in the evaluation of $\alpha_D$. When $\alpha_D$ is obtained from Eq.~(\ref{eq-polar2}), which incorporates symmetry-energy coefficients predicted from the IVGDR spectra, a strong correlation ($r>0.80$) between $\alpha_DJ$ and $r_\text{skin}$ is observed for heavy nuclei ($A>150$). Representative results for a few randomly selected nuclei are shown in Fig.~\ref{plot-polar2}. When $\alpha_D$ is evaluated using Eq.~(\ref{eq-polar1}), the correlation becomes even stronger, with $r>0.90$, as illustrated in Fig.~\ref{plot-polar2-new}. These results further indicate an approximately linear dependence of $r_\text{skin}$ on the product $\alpha_DJ$, which can be expressed as
\begin{equation}\label{eq-skin}
10^{-2}\alpha_DJ=b_1(A)r_\text{skin}+b_2(A),
\end{equation}
where the mass-dependent coefficients $b_1(A)$ and $b_2(A)$ are extracted separately for the two expressions of $\alpha_D$. The extracted coefficients are displayed in Fig.~\ref{plot-polar3}. Interestingly, both $b_1(A)$ and $b_2(A)$ exhibit an approximately linear dependence on the mass number $A$, leading to the parameterizations
\begin{align}
b_1(A) &=
\begin{cases}
0.087A-1.176 & \text{for Eq.~(\ref{eq-polar2})},\\
0.122A+4.879 & \text{for Eq.~(\ref{eq-polar1})},
\end{cases}\\
b_2(A) &=
\begin{cases}
0.049A-2.928 & \text{for Eq.~(\ref{eq-polar2})},\\
0.019A-1.513 & \text{for Eq.~(\ref{eq-polar1})}.
\end{cases}
\end{align}

Equation~(\ref{eq-skin}) enables the prediction of $r_\text{skin}$ for heavy nuclei from known or measured values of the product $\alpha_DJ$. To benchmark this relation, we predict $r_\text{skin}$ for the $^{208}$Pb nucleus and compare the result with the measurements reported by the PREX collaboration~\cite{Abrahamyan:2012gp,PREX:2021umo}, as well as with the value extracted from the electric dipole response~\cite{Tamii:2011pv}. The required $\alpha_D$ value at $A=208$ is obtained from Fig.~\ref{plot-polar} by interpolation and subsequently multiplied by the corresponding $J$ value of each EDF. The resulting product $\alpha_DJ$ is then inserted into Eq.~(\ref{eq-skin}) to determine $r_\text{skin}$ for $^{208}$Pb.

\begin{figure}
    \centering
    \includegraphics[width=\linewidth]{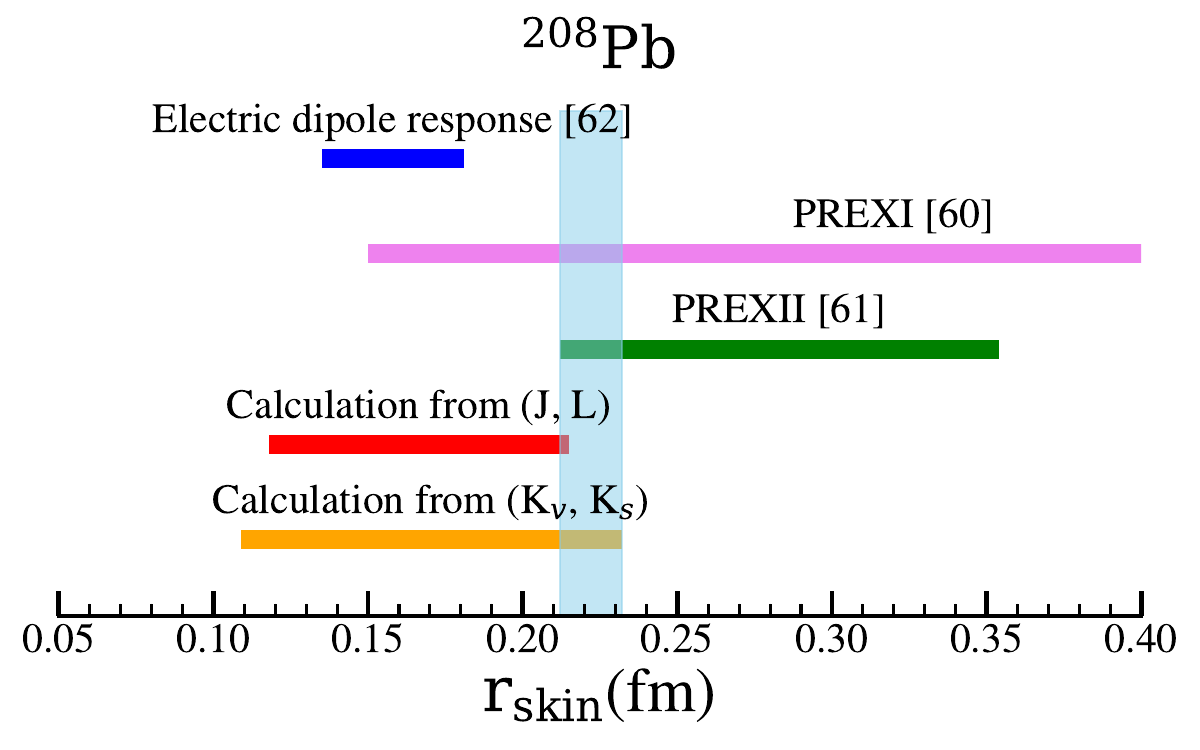}
    \caption{(Color online) Neutron skin thickness ($r_\text{skin}$) for $^{208}$Pb from our calculations and different measurements as indicated. The width of each band gives the uncertainty. The vertical shaded region gives the overlap between PREX measurements and the prediction from the analysis of IVGDR spectra.}
    \label{plot-polar4}
\end{figure}

The uncertainty ranges in $r_\text{skin}$ arising from the use of different EDFs are evaluated for both prescriptions of $\alpha_D$, namely Eq.~(\ref{eq-polar2}) and Eq.~(\ref{eq-polar1}), and the predictions are compared with the experimental data in Fig.~\ref{plot-polar4}. Both prescriptions yield a common range of $r_\text{skin}$ that overlaps with the measurements reported in~\cite{Tamii:2011pv} and with the lower part of the PREXI~\cite{Abrahamyan:2012gp} result. The uncertainties associated with the predictions based on IVGDR characteristics [Eq.~(\ref{eq-polar2})] are slightly larger than those obtained from Eq.~(\ref{eq-polar1}). Interestingly, however, the IVGDR-based predictions also overlap considerably with the range reported in the most recent PREXII~\cite{PREX:2021umo} measurement. Overall, these findings provide useful guidance for future experimental investigations and further demonstrate the applicability of Eq.~(\ref{eq-skin}) as a practical and efficient tool for estimating $r_\text{skin}$ in heavy nuclei.
%%%%%%%%%%%%%%%%%%%%%%  FIG 12 %%%%%%%%%%%%%%%%%%%%%%%%%%%%%%%%%%%%%%%%
%%%%%%%%%%%%%%%%%%%%%%%%%%%%%%%%%%%%%%%%%%%%%%%%%%%%%%%%%%%%%%%%%%%%%%%%%%%
\section{Conclusion}\label{sec:conclusion}
By applying Bayesian inference to the measured isovector giant dipole resonance (IVGDR) data, we have derived compact analytical parametrizations for both the centroid energy and the width of the IVGDR. Importantly, the proposed parametrizations explicitly incorporate the effects of nuclear isospin asymmetry and shell structure. The inclusion of isospin dependence allows the formulas to capture the evolution of collective dipole dynamics with neutron–proton imbalance, while the shell corrections account for local fluctuations arising from closed-shell configurations and structural effects. For practical applications, it is therefore possible to obtain a consistent description of resonance properties by using available nuclear structure information, such as deformations and shell corrections. Moreover, as the comparison with microscopic calculations suggests, the prescribed analytical form for the centroid energy is robust and globally applicable to stable nuclei.

These parametrizations are expected to be valuable for large-scale applications in nuclear astrophysics, where reliable photon strength functions are essential for calculating photon-induced and radiative capture reaction rates. These reactions typically occur at energies below the particle-emission threshold, where the low-energy dipole strength, including possible contributions from the pygmy dipole resonance (PDR), plays an important role. Unfortunately, systematic and consistent experimental information on the PDR remains scarce for a global analysis. Moreover, as discussed in Ref.~\cite{Goriely:1998utv}, the low-energy tail of the IVGDR strength function can also make a significant contribution to the reaction rates in this energy region. Furthermore, although characteristic astrophysical temperatures correspond to relatively low average photon energies, the high-energy tail of the thermal photon distribution gives rise to a small but finite population of energetic photons. In astrophysical environments with extremely high photon and nuclear abundances, these photons can excite giant resonances and induce photodisintegration reactions. In this context, the present analytical expressions provide an efficient, physically motivated, and computationally convenient description of the IVGDR strength function, making them well suited as input for large-scale astrophysical reaction-network calculations.

The symmetry energy coefficients extracted within this framework are found to be consistent with experimentally measured electric dipole polarizabilities. The dipole polarizability is known to be strongly correlated with the symmetry energy and its density dependence, thereby providing an important constraint on the nuclear equation of state. The excellent agreement obtained here demonstrates the internal consistency of our Bayesian analysis and strengthens the reliability of the inferred symmetry energy parameters.

An immediate implication of these results is the possibility of constructing an effective and predictive formulation for the neutron-skin thickness. Since the neutron-skin thickness is closely linked to both the dipole polarizability and the symmetry energy slope parameter, the present approach offers a practical pathway to estimate neutron skins across a broad mass range. Such a formulation will be highly valuable for guiding and interpreting future precision measurements of neutron-skin thickness, thereby contributing to improved constraints on the density dependence of the symmetry energy and its astrophysical implications, including neutron star structure.

%%%%%%%%%%%%%%%%%%%%%%%%%%%%%%%%%%%%%%%%%%%%%%%%%%%%%%%%%%%%%%%%%%%%%%%%%%%
\section*{Acknowledgments}
The authors also acknowledge Saha Institute of Nuclear Physics for providing access to the CCS3 high-performance computing cluster used for the calculations in this work. B.K.A. acknowledges with gratitude the support provided by the Department of Atomic Energy (DAE), Government of India, under the Raja Ramanna Chair Scheme.

\bibliographystyle{apsrev4-2}
\bibliography{reference_new}
\end{document}